\documentclass[aps,prd,showpacs,notitlepage,nofootinbib,superscriptaddress,floatfix,showkeys,twocolumn]{revtex4-1}
\usepackage{float}
\usepackage{siunitx}
\usepackage{verbatim}
\usepackage{graphicx}
\usepackage{amsmath}
\usepackage[usenames]{color}
\usepackage{amssymb}
\usepackage{hyperref}
\newcommand{\be}{\begin{equation}}
\newcommand{\ee}{\end{equation}}
\newcommand{\bea}{\begin{aligned}}
\newcommand{\eea}{\end{aligned}}
\newcommand{\pr}{\partial}

\newcommand{\bse}{\begin{subequations}}
\newcommand{\ese}{\end{subequations}}

\newcommand{\rb}{\bf r}
\newcommand{\xb}{\bf x}

\renewcommand{\v}[1]{\ensuremath{\mathbf{#1}}} 
\newcommand{\bmm}{\begin{multline}}
\newcommand{\emm}{\end{multline}}

\newcommand{\mi}{\mathrm{i}}
\begin{document}
\title{ A comparative study of the absorption cross section of static regular black holes for electromagnetic field}
\author{Rajesh Karmakar}
\email{rajesh018@iitg.ac.in}
\affiliation{Department of Physics, 
Indian Institute of Technology Guwahati, Assam 781039, India}

\begin{abstract}
In this article, we report our study on the absorption cross section for the electromagnetic field due to static spherically symmetric regular black holes of two distinct categories: the Fan-Wang generic class and the Simpson-Visser class. Existing studies on these two categories have relied on the WKB method, which applies to the low-frequency regime. In our analysis, we have numerically evaluated the absorption cross section for the electromagnetic field across all frequency ranges. To compute the absorption cross section, we followed a gauge-invariant approach, which avoids the need for gauge fixing for the electromagnetic field, thereby simplifying the methodology further. To emphasize the usefulness of this gauge invariant approach we have performed the same analysis for Simpson-Visser class black holes, which is more general in terms of the structure of the metric. For both types of regular black holes, the influence of regularity parameters on the absorption cross section reveals fundamental properties of their spacetime and their interactions with the surrounding astrophysical environment. Moreover, certain similarities observed between the Reissner-Nordstr\"om black hole and regular black holes of the Fan-Wang generic class indicate potential scenarios of correspondence between linear and nonlinear electrodynamics when coupled with gravity.

\end{abstract}

\maketitle

\newpage
\section{Introduction}
In the last decade, remarkable progress in the observation of various predictions of Einstein's general relativity (GR) has led to renewed interest in the field of black hole (BH) physics. Detection of gravitational waves (GWs) by LIGO-Virgo-KAGRA (LVK) \cite{LIGOScientific:2016aoc, LIGOScientific:2016emj, LIGOScientific:2016vbw, LIGOScientific:2017ycc, LIGOScientific:2017vwq, LIGOScientific:2020stg, LIGOScientific:2020aai, LIGOScientific:2020zkf, LIGOScientific:2023fpk, KAGRA:2023pio, KAGRA:2022twx} has created a new branch of physics called GW astronomy. The EHT collaboration's success in capturing the shadow images of Sgr A* and M87*  \cite{EventHorizonTelescope:2019dse, EventHorizonTelescope:2022wkp} provides direct evidence for the existence of BHs. These achievements in the observational front have spurred numerous studies on the premise of strong field gravity with various interesting outcomes \cite{Barack:2018yly, Perkins:2020tra, Carullo:2018sfu, Konoplya:2016pmh, Johannsen:2016vqy, Johannsen:2016uoh, Johannsen:2015hib, Vagnozzi:2022moj}. Still, fundamental challenges remain, including the unification of gravity and quantum mechanics, the singularity problem \cite{Hawking:1973, Joshi:2011rlc}, and others. One is left with the speculation as to whether a successful theory of quantum gravity could circumvent the issue of singularity. On the other hand, its appearance is inevitable from cosmic censorship conjecture due to Hawking and Penrose \cite{Hawking:1973, Joshi:2011rlc, Penrose:1969pc, Hawking:1979ig, Wald:1997wa}. Following this conjecture, the singularity is hidden inside the BH and dressed by the event horizon, therefore, its presence remains invisible. This brings us to the question of the naked singularity, which may form via the gravitational collapse of fundamental fields. However, the formation mechanism is much debated \cite{Christodoulou:1994hg1999, Singh:1997wa, Christodoulou:1994hg} and its observational premise is currently being intensively researched in light of the EHT findings \cite{ Deliyski:2024wmt, Vagnozzi:2022moj, Broderick:2024vjp}. In contrast, the proposals of regular BH (RBH) solutions are very promising as they are devoid of singularity. Therefore, the properties of these BHs need to be investigated rigorously.

RBHs have drawn much attention in recent times \cite{Riaz:2022rlx, Villani:2021lmo, Allahyari:2019jqz, Carballo-Rubio:2018pmi, Abdujabbarov:2016hnw}. The presence of the unphysical curvature singularity is cured in these BHs via coupling to the matter or through the modifications of Einstein GR. Such as the well known coupling of nonlinear electromagnetic fields with Einstein's GR leading to RBH of various categories: Bardeen class \cite{Bardeen1968}, Hayward class \cite{Hayward:2005gi} and so on \cite{Ayon-Beato:1998hmi}. Recently, Fan-Wang proposed a generic class of RBHs \cite{Fan:2016hvf}, which includes the previous two types along with the solutions for a new category of RBHs. In another scenario,  Simpson-Visser \cite{Simpson:2018tsi} constructed a special class of RBH solution through a coordinate transformation to the metric of Schwarzschild BH. This is a special class RBH in the sense that it also allows for a wormhole solution in certain parameter spaces. However, in our analysis, we will restrict ourselves to the parameter space generating, particularly, the RBH solution.

In this paper, we have studied the scattering of the EM field in the background of static spherically symmetric RBHs of two distinct classes: the Fan-Wang generic class \cite{Fan:2016hvf} and the Simpson-Visser class \cite{Simpson:2018tsi}, and evaluated the absorption cross section (ACS) for all the frequency ranges of the EM field by considering minimal interaction with the gravitational background. The scattering of various fundamental fields in the BH background has been the subject of active research area for a long time with various interesting outcomes \cite{Paula:2020yfr, Leite:2018mon, Teukolsky:1973ha, Teukolsky:1974yv, Unruh:1976fm, Fabbri:1975sa}. Existing studies on scalar fields within the context of Bardeen class \cite{Macedo:2014uga}, Hayward \cite{dePaula:2023muc} and Simpson-Visser class \cite{Lima:2020auu} RBHs have demonstrated the impact of the associated charge and regularity parameters of these spacetimes on the ACS. However, for these RBHs, the case of the EM field has not been explored more rigorously as compared to the scalar field scenario. The primary reason may be associated with the gauge ambiguities and nontriviality of the computation, specifically in the case of the Simpson-Visser class. Few studies that attempted to evaluate the ACS of RBHs for the EM field are as follows. For the Fan-Wang generic class, ACS has been evaluated with third-order WKB analysis in the low-frequency regime \cite{Lopez:2022uie}. Analysis for Bardeen \cite{Konoplya:2023ahd} and Bardeen-DeSitter \cite{Dey:2018cws} RBHs have been done, again using the WKB method. Using approximate method \cite{Visser:1998ke}, lower bounds of the greybody factor of the Simpson-Visser class have been studied \cite{Jha:2023wzo}. In our analysis, we have extended some of these studies to the high-frequency range, specifically in the case of the Fan-wang generic class and compared them with the existing results. Concerning the derivation of the EM field equation and subsequent analysis, there have been two approaches: one by fixing the gauge \cite{Crispino:2007qw, Crispino:2008zz, deOliveira:2019tlk} and another by following a gauge invariant formalism \cite{Teukolsky:1973ha, Teukolsky:1974yv}. The latter approach also involves the Kinnersley null tetrad formalism, specifically used to study the axis-symmetric BH spacetime and to derive the Teukolsky equation \cite{Teukolsky:1973ha, Teukolsky:1974yv}. In the present analysis, we have followed a gauge invariant methodology, which seems to be much simpler than the null tetrad formulations. This method appeared only once in one of the author's earlier works \cite{Karmakar:2023hlb}. We find this approach to be effective in computing the ACS of RBHs, specifically for the Simpson-Visser class.

The rest of the paper has been organized as follows. In Sec.\ref{diff.RBHs}, we begin with a brief overview of the two classes of RBHs and their characteristics. Next, in Sec.\ref{coupling} we have formulated the minimally coupled EM field equations following the gauge invariant approach. This section also illustrates the properties of the effective scattering potential for the EM field in the respective background of RBHs. After that, in Sec.\ref{method.ACS}, we have developed this methodology further to compute the ACS for the EM field. In Sec.\ref{numeric.analysis}, we have presented our numerical findings along with the explanations of various characteristic features of ACS for RBHs. In addition to that, we have discussed the possible scenario in which RBHs belonging to the Fan-Wang generic class can mimic the Reissner–Nordström BH. Finally, in Sec.\ref{concl}, we conclude with possible future directions.

Throughout the analysis, we will use the metric signature as $(-,+,+,+)$ and follow the natural units, $\hbar=c=G=1$.

\section{Regular black holes of two distinct classes}\label{diff.RBHs}
Generally, the line element for a static spherically symmetric RBH spacetime is written as,
\be\label{gen.rbh}
ds^2=-f(r)dt^2+p(r)^{-1}dr^2+r^2d\theta^2+r^2\sin^2\theta d\phi^2,
\ee
where $f(r)$ and $p(r)$ are the generic lapse functions. This metric allows for a time-like killing vector field that is non-vanishing and hypersurface orthogonal \cite{Padmanabhan:2010zzb}. The general definition of the tortoise coordinate for the above spacetime can be given as $dr_*=dr/\sqrt{fp}$. In the following discussion, we have given a brief overview of the two distinct classes of RBHs that will be considered in our analysis. 
\subsection{Fan-Wang generic class}
The standard GR theory coupled with nonlinear electrodynamics (NLED) is described by the following action \cite{Fan:2016hvf},
\be
S=\int d^4x \sqrt{-g}\left[\frac{1}{26\pi}R-\frac{1}{4\pi}\mathcal{L}(F)\right],
\ee
where $R$ denotes the Ricci scalar and the Lagrangian of the NLED symbolized by $\mathcal{L}$ is a function of the EM field strength, $\mathcal{F}=F_{\mu\nu}F^{\mu\nu}$.  A generic solution of RBHs arising from the coupling of gravity to the NLED has been proposed by Fan-Wang \cite{Fan:2016hvf}, where the NLED lagrangian is considered as (also see \cite{dePaula:2023muc}), 
\be\label{mat.Lagrangian}
\mathcal{L}=\frac{4M\mu}{q^3}\frac{\left(q^2 \mathcal{F}\right)^{\frac{\nu+3}{4}}}{\left(1+\left(q^2\mathcal{F}\right)^{\frac{\nu}{4}}\right)^{\frac{\mu+\nu}{\nu}}}.
\ee
The regular BH solution sourced by this Lagrangian can be expressed as \eqref{gen.rbh}, with the following lapse functions,
\be
f(r)=p(r)=1-\frac{2Mr^{\mu-1}}{(r^\nu+q^\nu)^{\mu/\nu}},
\ee
where $M$ represents the ADM mass \cite{Ayon-Beato:1998hmi}. Whereas, $q$ denotes the magnetic charge (in some cases it may represent the electric charge also but we will not consider that here). This RBH solution essentially describes the spacetimes of magnetically charged BHs, often interpreted as being sourced by a nonlinear magnetic monopole \cite{Ayon-Beato:2000mjt}. To get rid of the singularity at $r\to 0$, the BH parameter, $\mu$ is restricted as $\mu\geq 3$. However, we will fix $\mu=3$ for the sake of simplicity in our discussion. Then, the other parameter $\nu>0$, determines various categories of RBHs in the NLED theory. This set of RBH classes includes Bardeen class \cite{Bardeen1968} and Hayward class BHs \cite{Hayward:2005gi} among others, as has been presented in Table \ref{RBH.tab}. The Fan-wang generic class also provides for a new class of RBHs with $\mu=3$ and $\nu=1$. It is worth mentioning that in the weak field limit, the behaviour of Bardeen class, $\mathcal{L}\sim \mathcal{F}^{5/4}$, Hayward class, $\mathcal{L}\sim \mathcal{F}^{3/2}$ and for the new class, $\mathcal{L}\sim \mathcal{F}$, can be inferred from \eqref{mat.Lagrangian}. Therefore, the new class RBHs approaches the standard Maxwell theory, while the first two classes are comparatively stronger. Also, among them, the Bardeen class is closer to the Maxwell theory than the Hayward class. These correspondences with linear theory will be evident in our final results, as will be shown in the later sections.

For such RBHs we have the restriction $q^\nu\leq q^\nu_{crit}$, otherwise, this will lead to an extremal condition \cite{Macedo:2014uga, dePaula:2023muc, Lopez:2022uie}. The critical value of the magnetic charge, $q^\nu_{crit}$ can be derived by the extremization condition, $dq^\nu(r_h)/dr_h=0$ at some $r_h=r_{crit}$ with $r_h$ representing the roots of $f(r_h)=0$. This finally leads to the following expression of the critical value of the charge \cite{Lopez:2022uie},
\be
q_{crit}\equiv q(r_{crit})=2M\left(\frac{\mu}{\mu-1}\right)^{-\mu/\nu}\left(\frac{1}{\mu-1}\right)^{1/\nu} 
\ee
We have given the values of the critical charges for respective categories of RBHs in Table \ref{RBH.tab}. Nevertheless, we will investigate the effect of such restriction on charges in the ACS for the EM field, with a particular focus on the new class RBH.
\renewcommand{\arraystretch}{1.6}
\begin{table}[t]
 \centering
\begin{tabular}{|c|c|c|} 
\hline
 Parameter & Nomenclature & Critical charge \\
 ($\mu$, $\nu$) &  & $q_{crit}\simeq$ \\
 \hline
 \hline
 ($3$, $1$)& Newc & $0.2963 M$\\
 \hline
 ($3$, $2$)& Barc & $0.7698 M$\\
\hline
 ($3$, $3$)& Hayc & $1.0583 M$\\
\hline
\end{tabular}
\caption{Various classes of RBHs for different sets of RBH parameters with associated critical charges. We have used the abbreviation for Newclass as Newc, Bardeen class as Barc, and Hayward class as Hayc. This nomenclature will also be used in all the plots presented in the following sections.}\label{RBH.tab}
\end{table}
\subsection{Simpson-Visser class}
 Simpson-Visser proposed an RBH solution \cite{Simpson:2018tsi} by applying a coordinate transformation to the line element of the vacuum Schwarzschild BH. In this sense, this particular class is drastically different from the RBH arising from the NLED theory. Recently, however, it has been shown \cite{Bronnikov:2021uta} that these RBHs may emerge from the coupling of gravity to a combination of the phantom-like scalar field, and NLED Lagrangian of the form discussed before. Moreover, RBHs of this class have an interesting feature of connecting the Schwarzschild BH to the Morris-Thorne traversable wormhole. Even in the regime where it describes RBHs, this solution exhibits unusual properties, as it represents spacetimes with a \textit {one-way space like throat} shielded by a single horizon \cite{Simpson:2018tsi}. The line element for this class of solution can be cast (see \cite{Franzin:2023slm}) in the form of \eqref{gen.rbh} with the following lapse functions,
\be
f(r)=\left(1-\frac{r^2_0}{r^2}\right)^{-1}\left(1-\frac{2M}{r}\right),~~p(r)=\left(1-\frac{2M}{r}\right),
\ee
where $r_0$ is the regularization parameter. The value of this parameter determines the characteristics of the space-time as to whether it is a BH, RBH or wormhole solution. For the present study, we are specifically interested in the RBHs, which are obtained with $r_0<2M$, accordingly, we will restrict the value of this parameter in our analysis.  Also, note that $M$ here signifies the Schwarzschild mass.

The above two categories largely comprise the generic existing classes of RBHs \cite{Carballo-Rubio:2019fnb}. In the NLED context, magnetic charge, $q$ plays the role of regularization, whereas, the same role is played by $r_0$ in the context of the Simpson-Visser class. Our primary motivation is to study how these parameters impact the ACS for the EM field. 
\begin{figure*}[t]
\includegraphics[width=\linewidth]{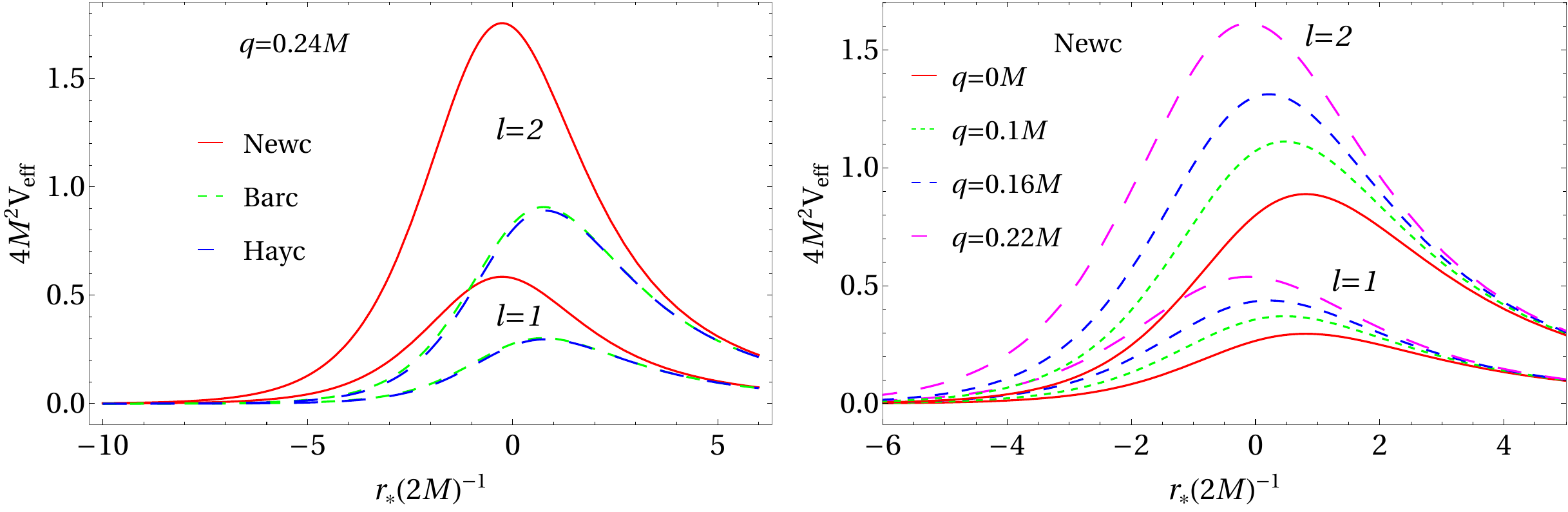}
\caption{Potential barrier for the EM field with $r_*$: In the left panel, we have considered, Fan Wang generic class RBHs for a fixed charge $q=0.24M$. In the right panel, we have considered only the new class among the Fan Wang class RBHs for different charges within the critical charge, $q_{crit}\simeq 0.2963 M$, also the red line, for $q=0$, represents the case for Schwarschlid BH. Variations have been shown for two orbital modes $l=1$ and $l=2$.}\label{NBHandNVeffl1l2}
\end{figure*}
\section{Minimally Coupled Gauge Field and partial wave analysis}\label{coupling} 
The governing equation of motion for a minimally coupled EM field is 
\be\label{Eom}
\bea
& \frac{1}{\sqrt{-g}}\pr_\mu(\sqrt{-g}g^{\mu\alpha}g^{\nu\beta}F_{\alpha\beta})=0 .
\eea
\ee
For a general spherically symmetric spacetime one can decompose the components of the EM field in the following manner \cite{Brito:2015oca}
\be\label{modedecom},
\bea
&A_t(t,\v r)=\sum_{lm} b^{lm}(t,r)Y_{lm}(\theta,\phi) ,\\
&A_r(t,\v r)=\sum_{lm}h^{lm}(t,r)Y_{lm}(\theta, \phi) ,\\
&A_{s}(t,{\rb})=\sum_{lm}\left[k_{lm}(t,r)\Psi^{lm}_{s}(\theta, \phi)+a_{lm}(t,r)\Phi^{lm}_{s}(\theta,\phi)\right],
\eea
\ee
by using the orthonormal vector spherical harmonic basis \cite{Barrera, Jackson:1998nia} to write the spherical part of the field components. Here, $b^{lm}, h^{lm}, k^{lm}$ and $a^{lm}$ are some regular function of time and spatial coordinate. Whereas, expression of the spherical harmonics are \cite{Barrera, Brito:2015oca},
\be
\bea
&\Psi^{lm}_s=\pr_s{Y_{lm}},\\
&\Phi^{lm}_s=\epsilon_{ss'}\pr^{s'}{Y_{lm}} .
\eea
\ee
Where, $s,s'$ indices correspond to angular coordinates $(\theta,\phi)$ and $\epsilon_{\theta\theta}=\epsilon_{\phi\phi}=0,\epsilon_{\theta\phi}=-\epsilon_{\phi\theta}=\sin\theta$. The above decomposition will help us to separate the spherical part of the gauge field that we will discuss next. One can see that the decomposed fields under gauge transformation will lead to the following combinations \cite{Brito:2015oca}, which are gauge invariant, 
\be\label{invariant.var}
\bea
&\chi^{lm}_1=\frac{r^2}{l(l+1)}(\pr_t h^{lm}-\pr_r b^{lm}), \\
&\chi^{lm}_2=a^{lm}, \\
&\chi^{lm}_3=h^{lm}-\pr_r k^{lm},\\
&\chi^{lm}_4=b^{lm}-\pr_t k^{lm} .
\eea
\ee
In our subsequent analysis, we work with these gauge-invariant variables. Substituting the field components from \eqref{modedecom} to the equation of motion \eqref{Eom} and expressing it in terms of the gauge invariant variables \eqref{invariant.var}, the field equation can be separated from the spherical part and can be expressed in the following manner, 
\be\label{chi1.eom}
\bea
&f(r)\pr_r\left[\sqrt{f(r)p(r)}~\pr_r\left(\sqrt{\frac{p(r)}{f(r)}}\chi^{lm}_1\right)\right]\\
&~~~~~~~~~~~~~~~~~~-\pr^2_t\chi^{lm}_1-f(r)\frac{l(l+1)}{r^2}\chi^{lm}_1=0.
\eea
\ee
\begin{figure}[t]
\includegraphics[scale=0.4]{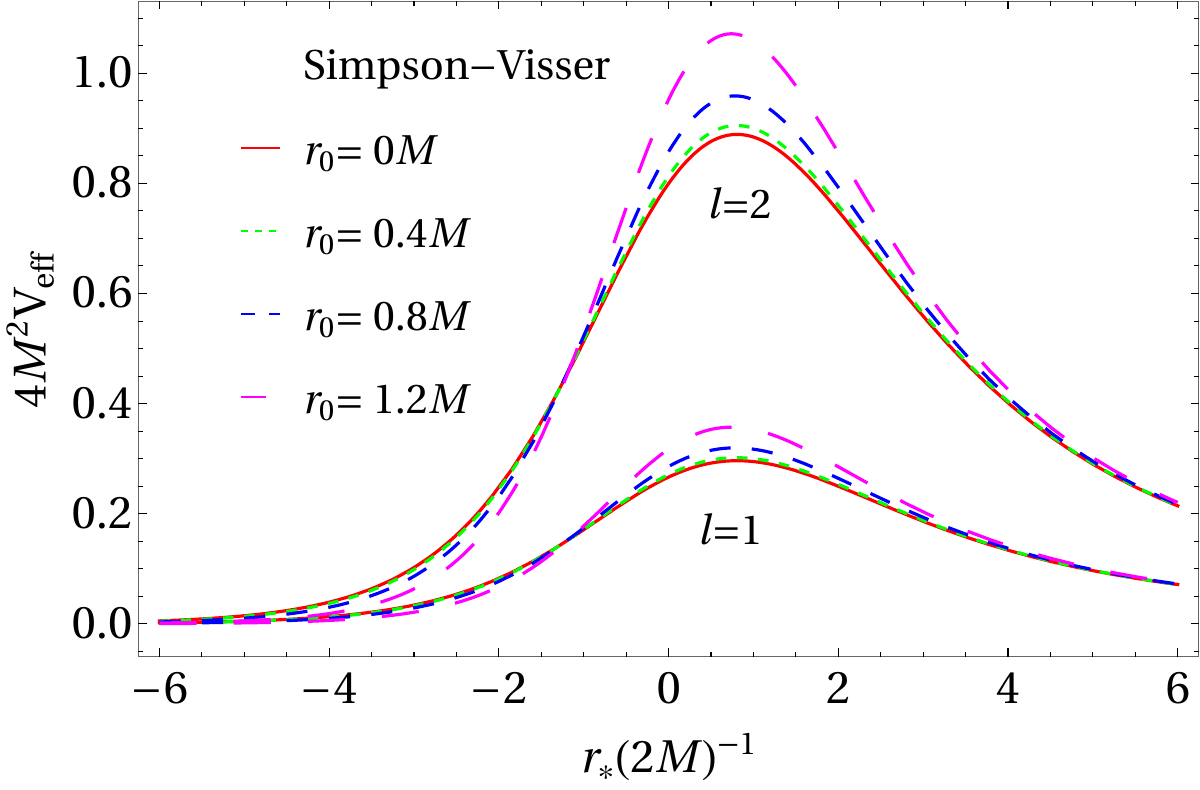}
\caption{Potential barrier for the EM field in Simpson-Visser BH spacetime. Variation is shown by varying BH regularity parameter $r_0$ for the orbital mode $l=1$ and $l=2$. Important to note that the red line for $r_0=0$, represents the case for Schwarschlid BH.}\label{SV_Veff}
\end{figure}
Similarly, we obtain the governing equation of the $\chi_2$ as,
\be\label{chi2.eom}
\bea
&\sqrt{f(r)p(r)}\pr_r\left(\sqrt{f(r)p(r)}\pr_r \chi^{lm}_2\right)\\
&-\pr^2_t \chi^{lm}_2-f(r)\frac{l(l+1)}{r^2}\chi^{lm}_2=0.
\eea
\ee
As it turns out the other two gauge invariant variables are not independent and are constrained as,
\be\label{eomchi34}
\bea
&f(r)\pr_r\chi^{lm}_1+\chi^{lm}_4=0,\\
&\pr_t \chi^{lm}_1+f(r)\chi^{lm}_3=0 .
\eea
\ee
Utilizing the static nature of the RBHs, we re-express the two independent gauge invariant variables in the following form,
\be\label{tr.decomp}
\bea
\chi^{lm}_1(t,r)&\equiv e^{-\mi\omega t}\sqrt{f(r)/p(r)}\bar{\chi}^{lm}_1(r),\\
\chi^{lm}_2(t,r)&\equiv e^{-\mi\omega t}\bar{\chi}^{lm}_2(r).
\eea
\ee
Then the equations of motion \eqref{chi1.eom} and \eqref{chi2.eom} can be recast into the following time-independent form, 
\be\label{chi1chi2.eom}
\bea
&\pr^2_{r_*}\bar{\chi}^{lm}_1+\left[\omega^2-V_{eff}(r)\right]\bar{\chi}^{lm}_1(r)=0,\\
&\pr^2_{r_*}\bar{\chi}^{lm}_2+\left[\omega^2-V_{eff}(r)\right]\bar{\chi}^{lm}_2(r)=0,\\
\eea
\ee
where the effective potential turns out to be $V_{eff}\equiv f(r)l(l+1)/r^2$, which we have plotted in Fig.\ref{NBHandNVeffl1l2} and Fig.\ref{SV_Veff} for the two types of RBHs. In the left panel of Fig.\ref{NBHandNVeffl1l2}, we illustrate the effective potential for RBHs in the Fan-Wang generic class. With a fixed charge, we observe that the peak of the effective potential is lower for RBHs with higher critical charges, $q_{crit}$. In the right panel of Fig.\ref{NBHandNVeffl1l2}, we specifically consider the effective potential for the New class RBH with different charges,  restricted within $q_{crit}$. Here, the peak increases with the charge $q$. Additionally, we have also provided the effective potential for a Schwarzschild BH (for $q=0$) for visual comparison. It is worth mentioning that very similar features have also been observed for a scalar field in the case of the Bardeen class \cite{Macedo:2014uga} and Hayward class \cite{dePaula:2023muc}. Coming to the Simpson-Visser class, the effective potential plotted in Fig.\ref{SV_Veff} shows an increasing behaviour with the regularity parameter $r_0$ starting from the case (for $r_0=0$) of Schwarzschild BH. Importantly, the nature of the effective potentials for both categories clearly shows that they approach zero near the event horizon of the RBH and diminish very rapidly as they approach spatial infinity due to the $1/r^2$ factor. This particular characteristic is the same as that of the usual vacuum BH solutions due to the presence of an event horizon and it allows for the following asymptotic solutions, suitable to define the ACS, 
\be\label{bc.chi1chi2}
\bar{\chi}^{lm}_\lambda\to \begin{cases} \mathcal{N}^{\omega l}_\lambda\mathcal{T}^{\omega l}_\lambda e^{-\mi\omega r_* }, ~~~~~~~~~~~~~~~~~~~~~r_*\to -\infty,\\
\mathcal{N}^{\omega l}_\lambda\left(\mathcal{I}^{\omega l}_\lambda e^{-\mi\omega r_*}+\mathcal{R}^{\omega l}_\lambda e^{\mi\omega r_*}\right), ~~~~~~~r_*\to\infty,
\end{cases}
\ee
where $\lambda\equiv(1,2)$. Whereas, $\mathcal{T}^{\omega l}_\lambda, \mathcal{I}^{\omega l}_\lambda$ and $\mathcal{R}^{\omega l}_\lambda$ represent the transmission, incidence and reflection coefficients respectively. The normalization constants, $\mathcal{N}^{\omega l}_\lambda$ will be fixed in the following section. Next, we will develop the methodology to compute the ACS for the EM field utilizing the gauge invariant variables \eqref{invariant.var} defined in this section.
\section{Quantifying the absorption cross section (ACS)}\label{method.ACS}
We begin with a circularly polarized ingoing plane EM wave along $z$-direction, which can be expressed as \cite{Jackson:1998nia}, 
\be
\bea
&A_x(t,{\bf x})=e^{-\mi\omega(t+z)},\\
&A_y(t,{\bf x})=\mi e^{-\mi\omega(t+z)} .
\eea
\ee
The method to determine the normalization factor involves comparing the incident plane wave with the asymptotic form of the field solution, which is derived in spherical coordinates. 
Therefore, we need to transform the plane electromagnetic wave from Cartesian coordinates, $\textbf{A}=A_{x}\hat{x}+A_{y}\hat{y}$, to spherical coordinates, utilizing the transformation rule \cite{Padmanabhan:2010zzb}, 
\be
A'_{\mu}(x')=\frac{\partial x^{\nu}}{\partial x'^{\mu}}A_{\nu}(x),
\ee
so that the components of the Em field become, 
\be\label{pln.sph1}
\bea
&A'_t(t,\textbf{r})=A_t(t,\textbf{x}),\\
&A'_r(t,\textbf{r})=\sin\theta e^{\mi\phi}A_x(t,{\xb}),\\
&A'_\theta(t,\textbf{r})=r\cos\theta e^{\mi\phi}A_x(t,{\xb}), \\
&A'_\phi(t,\textbf{r})=ir\sin\theta e^{\mi\phi}A_x(t,{\xb}).
\eea
\ee
\begin{figure*}[t]
\includegraphics[width=\linewidth]{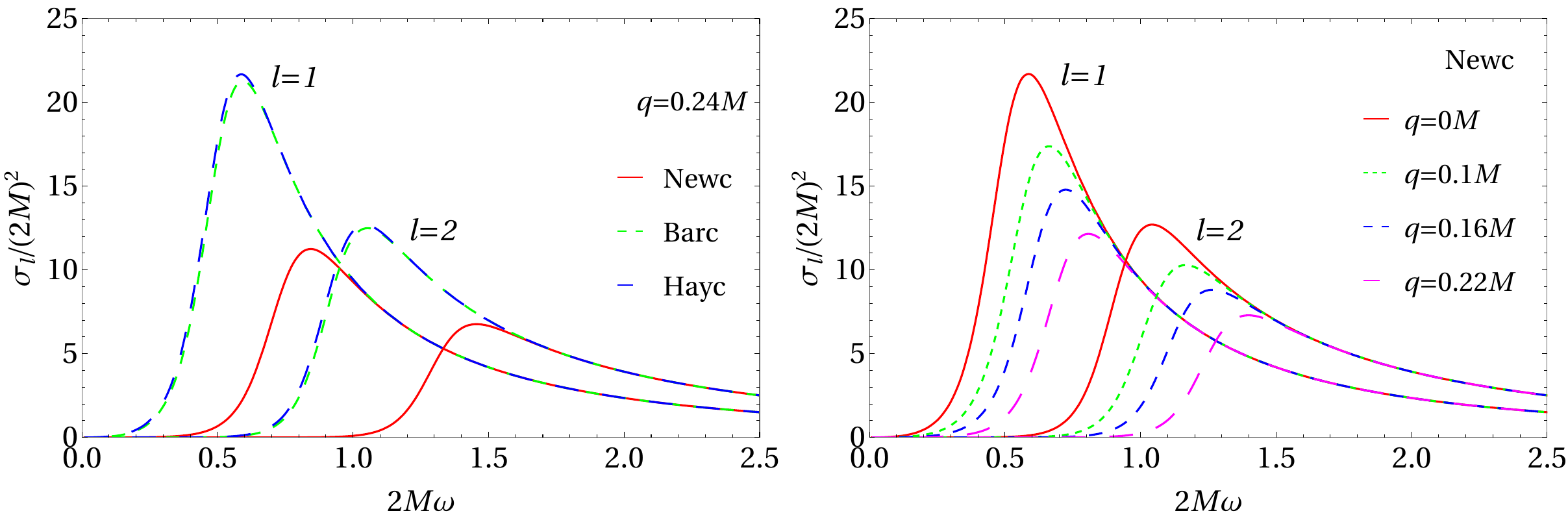}
\caption{Patial ACS for the EM field with frequency ($\omega$): In the left panel, we have considered, Fan Wang generic class RBHs for a fixed charge $q=0.24M$. In the right panel, we have considered, only the new class RBHs for different charges within the critical charge, $q_{crit}\simeq 0.2963 M$,  also the red line for $q=0$, represents the case for the Schwarschlid BH. Variations have been shown for two orbital modes $l=1$ and $l=2$.}\label{NBHandNpartACS}
\end{figure*}
\begin{figure}[t]
\includegraphics[scale=0.43]{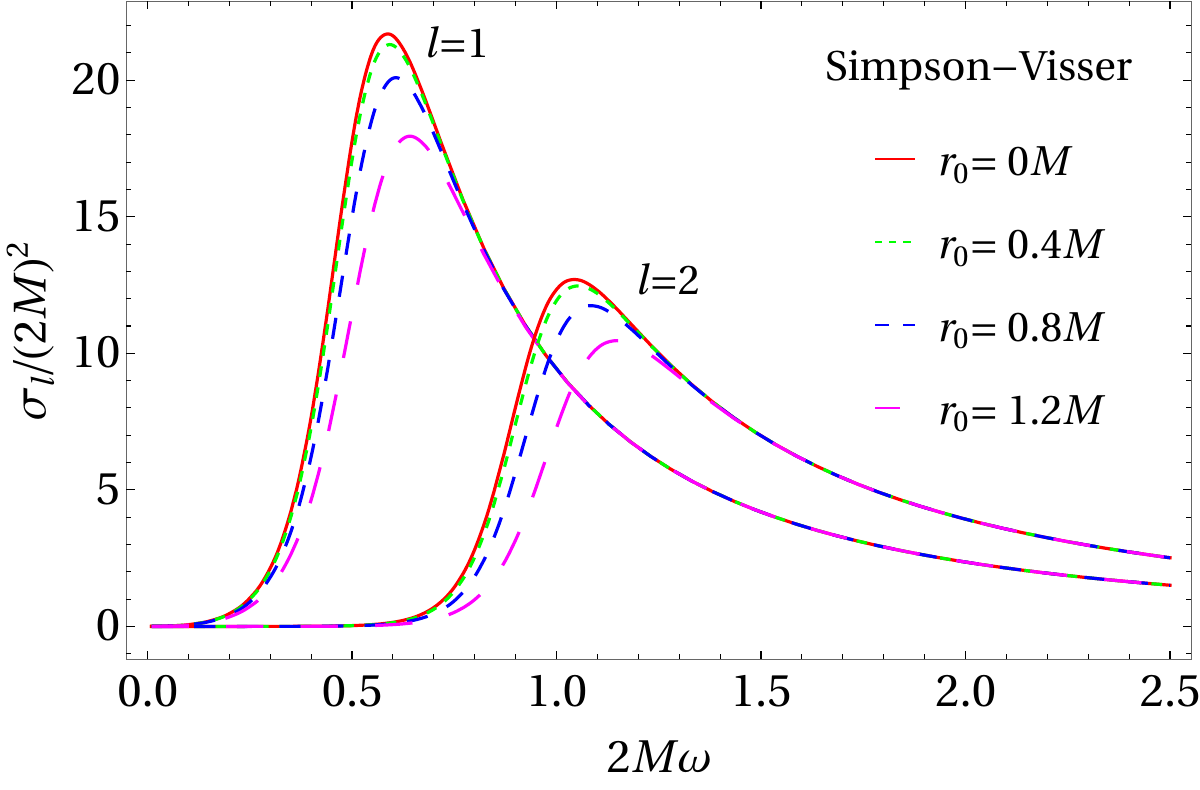}
\caption{Partial ACS of Simpson-Visser RBH for the EM field with frequency ($\omega$) has been presented considering different values of RBH parameter $r_0$. Important to note that the red line for $r_0=0$, represents the case for Schwarschlid BH.}\label{SV.partial.ACS}
\end{figure}
In the above analysis, we have also used $A_{z}=0$ and $A_y=\mi A_x$, the incident EM wave being circularly polarized. Now, $A_x(t,x)$ appearing on the right-hand side of the above decomposition should be expressed in spherical coordinates. To achieve this, we use the Rayleigh expansion of the spatial part of a plane wave propagating along the z-direction \cite{Griffiths.2018},
\be
e^{-\mi\omega z}=e^{-\mi\omega r\cos\theta}=\sum_{l=0}(2l+1)\mi^l j_{l}(\omega r)P^0_l(\cos\theta),
\ee
where $j_l(\omega r)$ is the spherical Bessel function and $P^0_l(\cos\theta)$ is the associated Legendre's function \cite{Abramowitz:1968}. By taking the derivative with respect to $\theta$ on both sides of the above equation, the decomposition can be re-expressed as follows:
\be
e^{-\mi\omega z}=e^{-\mi\omega r\cos\theta}=\sum_{l=0}(2l+1)\mi^l \frac{j_{l}(\omega r)}{\mi\omega r}\frac{\pr_\theta P^0_l(\cos\theta)}{\sin\theta}.
\ee
Where one needs to utilize the relation for associated Legendre's functions, $P^1_l(\cos\theta)=-\pr_\theta P^0_l(\cos\theta)$. Finally, the components of the plane electromagnetic waves for circularly polarized light in spherical coordinates turn out as \cite{Crispino:2007qw},
\be\label{pln.sph2}
\bea
&A'_t(t,\textbf{r})=\sum_{lm} A'^{lm}_t(t,r) Y_{lm}(\theta,\phi)=0,\\
&A'_r(t,\textbf{r}),\\
&=\sum_{lm} A'^{lm}_r(t,r)Y_{lm}(\theta, \phi)\\
&=\sum_{lm}(-1)^{l+1}\delta_{m1}\sqrt{4\pi(2l+1)l(l+1)}\frac{e^{-\mi\omega(t+r_*)}}{2\omega^2r^2}Y_{lm}(\theta, \phi)\\
&~~~~~~~~~~~~+{\rm outgoing~part},\\
&A'_{s}(t,{\rb})=-\sum_{lm}(-1)^{l+1}\delta_{m1}\sqrt{\frac{4\pi(2l+1)}{l(l+1)}}\times\\
&\times\Big[\mi\Psi^{lm}_{s}(\theta, \phi)+\Phi^{lm}_{s}(\theta, \phi)\Big]\frac{e^{-\mi\omega(t+r_*)}}{2\omega}+{\rm outgoing~part}.\\
 \eea
 \ee
Note that, $s$ in the subscript denotes the components corresponding to $(\theta,\phi)$ as before. Whereas, the $\delta_{m1}$ factor in the above expressions arises due to having $e^{i \phi}$ in the $A'_t$ and $A'_{\phi}$ components of the gauge field (see Eq.\ref{pln.sph1}). The above expansions can be verified by using the following relations \cite{Abramowitz:1968}: 
\be
\bea
&j_{l}(\omega r)+\mi n_{l}(\omega r)\sim\frac{(-\mi)^{l+1}e^{\mi\omega r}}{\omega{r}},\\
&j_{l}(\omega r)-\mi n_{l}(\omega r)\sim\frac{\mi^{l+1} e^{-\mi\omega r}}{\omega{r}},
\eea
\ee
for $r\to\infty$ or $\omega r>>1$, that implies 
\be
j_l(\omega r)\sim i^{l+1}\frac{e^{-\mi\omega r}}{2\omega r}+(-i)^{l+1}\frac{e^{\mi\omega r}}{2\omega r}.
\ee
With the components of the EM field \eqref{pln.sph2}, the gauge-invariant variables \eqref{invariant.var} for circularly polarized light take the following form,  
\be\label{inwave}
\bea
&\sum_{lm}\chi^{lm}_1(t,r)Y^{lm}(\theta, \phi)\\
&=-\mi\sum_{lm}(-1)^{l+1}\delta_{m1}\sqrt{\frac{4\pi(2l+1)}{l(l+1)}}\frac{e^{-\mi\omega(t+r_*)}}{2\omega}Y_{lm}(\theta, \phi)\\
&~~~~~~~~~~~~~~~~~+{\rm outgoing~part}.\\
&\sum_{lm}\chi^{lm}_2(t,r)\Phi^{lm}_s(\theta, \phi)\\
&=-\sum_{lm}(-1)^{l+1}\delta_{m1}\sqrt{\frac{4\pi(2l+1)}{l(l+1)}}\frac{e^{-\mi\omega(t+r_*)}}{2\omega}\Phi^{lm}_{s}(\theta, \phi)\\
&~~~~~~~~~~+{\rm outgoing~part}\\
\eea
\ee
\begin{figure*}[t]
\includegraphics[width=\linewidth]{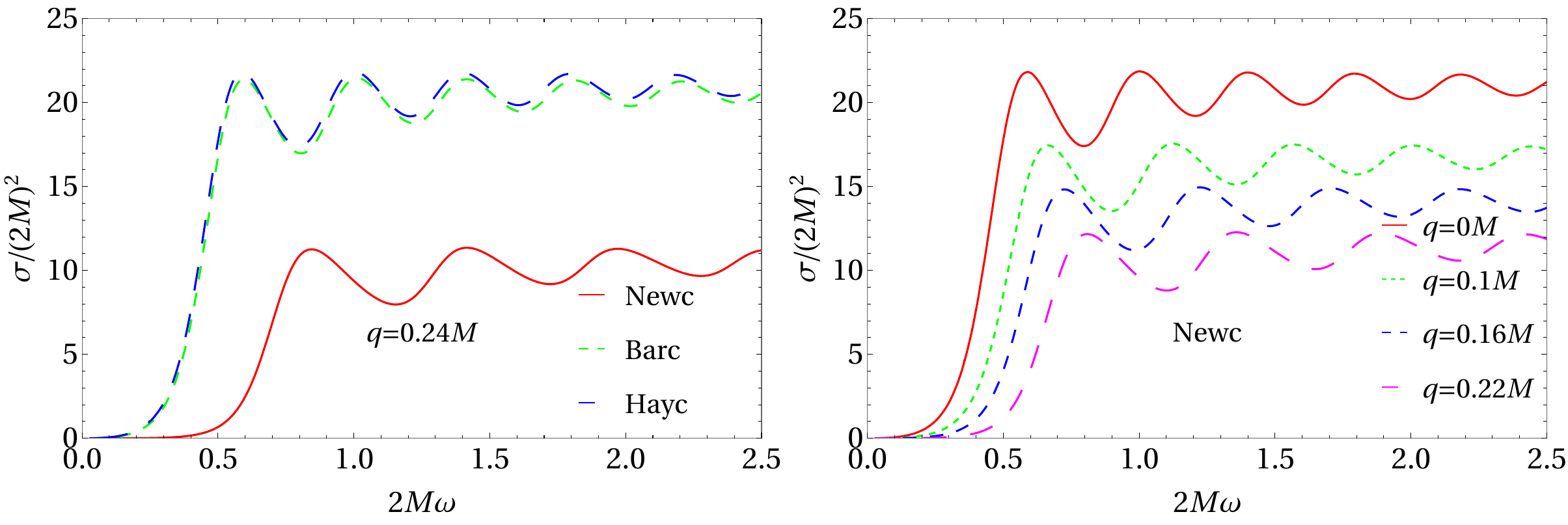}
\caption{Total ACS for the EM field with frequency ($\omega$): In the left panel, we have considered Fan Wang generic class RBHs for a fixed charge $q=0.24M$. In the right panel, we have considered, only new class RBHs for different charges within the critical charge, $q_{crit}\simeq 0.2963M$, also the red line for $q=0$ represents the case for Schwarschlid BH.}\label{NBHandNtotACS}
\end{figure*}
\begin{figure}[t]
\includegraphics[scale=0.43]{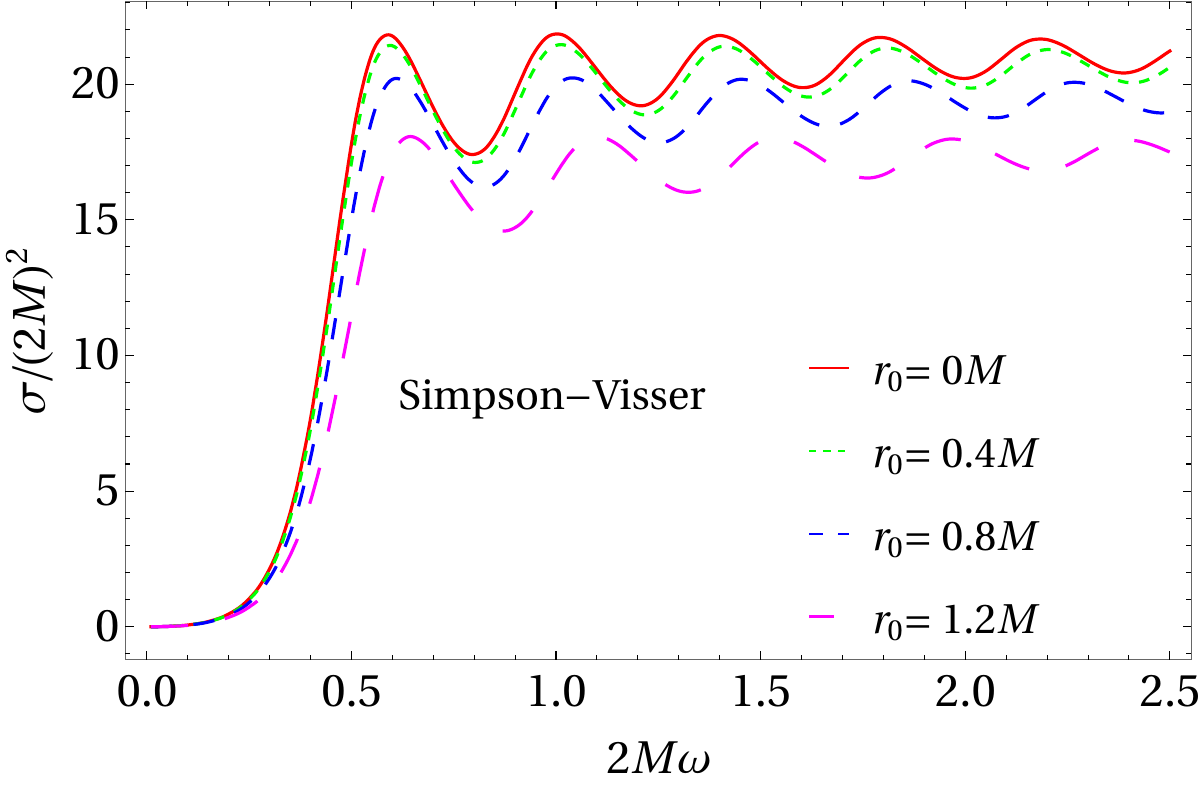}
\caption{Total ACS of Simpson-Visser RBH for the EM field with frequency ($\omega$) has been presented considering different values of RBH parameter $r_0$. Important to note that the red line for $r_0=0$, represents the case for Schwarschlid BH.}\label{SV_tot_acs_diffr0}
\end{figure}
Utilizing these expansions of the EM field, one can deduce the normalization factors from the asymptotic form of these fields, expressed in terms of the incidence and reflection amplitude \eqref{bc.chi1chi2}, 
\be\label{asympformchi1chi2}
\bea
&\sum_{lm}\chi^{lm}_1(t,r)Y^{lm}(\theta, \phi)\\
&=\sum_{lm}\mathcal{N}^{\omega l}_1[\mathcal{I}^{\omega l}_1e^{-\mi\omega(t+r_*)}+\mathcal{R}^{\omega l}_1e^{-\mi\omega (t-r_*)}]Y_{lm}(\theta, \phi), \\
&\sum_{lm}\chi^{lm}_2(t,r)\Phi^{lm}_s(\theta, \phi)\\
&=\sum_{lm}\mathcal{N}^{\omega l}_2[\mathcal{I}^{\omega l}_2e^{-\mi\omega(t+r_*)}+\mathcal{R}^{\omega l}_2e^{-\mi\omega(t-r_*)}]\Phi^{lm}_{s}(\theta, \phi).
\eea
\ee
Note that, here, we have omitted the metric coefficients, $f(r)$ and $p(r)$, which appeared in \eqref{tr.decomp} as we are dealing with the asymptotic behaviour in the limit $r\to \infty$. Comparing with \eqref{inwave} we fix the normalization factor \eqref{bc.chi1chi2} for $\bar{\chi}^{lm}_\lambda$ to be
\be\label{norm.fact}
\bea
&\mathcal{N}^{\omega l}_\lambda=-\mi(-1)^{l+1}\delta_{m1}\sqrt{\frac{4\pi(2l+1)}{l(l+1)}}\frac{1}{2\omega\mathcal{I}^{\omega l}_\lambda}.
\eea
\ee
With these normalization constants, one can suitably define the unit amplitude incident EM wave. Now, the ACS is defined \cite{Unruh:1976fm}  as the ratio of the energy absorbed by the BH horizon per unit time to the incident energy density. Furthermore, the part of the absorbed energy can be translated to the energy measured by a static observer at spatial infinity as,
\be
\pr_t\mathcal{E}=\int {r^2}d\Omega T_{tr}
\ee
where $d\Omega=\sin\theta d\theta d\phi$. In principle, this quantity estimates the energy entering a spherical surface considered at spatial infinity. Whereas the stress energy tensor of the EM field is \cite{Padmanabhan:2010zzb}
\be
T_{\alpha\beta}=-\frac{1}{4}g_{\alpha\beta}F_{\mu\nu}F^{\mu\nu}+g^{\nu\rho}F_{\alpha\nu}F_{\beta\rho}.\\
\ee
Substituting the EM field components \eqref{modedecom} in terms of the gauge invariant variables \eqref{invariant.var} in the above expression, one will be able to obtain,
\be
\bea
\pr_t\mathcal{E}&=\sum_{lm}\Big[\frac{l(l+1)}{2}\Big\{\Big(\pr_r\chi^{lm}_1\pr_t \chi^{lm^*}_1+\pr_t\chi^{lm}_2\pr_r\chi^{lm^*}_2\Big)\\
&~~~~~~~~~~~~~~~~~~~+\Big(\pr_t\chi^{lm}_1\pr_r \chi^{lm^*}_1+\pr_r\chi^{lm}_2\pr_t\chi^{lm^*}_2\Big)\Big\}\Big]
\eea
\ee
With the help of \eqref{asympformchi1chi2} along with the normalization factor \eqref{norm.fact} we arrive at the following formula,
\be
\pr_t\mathcal{E}==\sum_{l,\lambda}\pi (2l+1)\omega^2\left(1-\left|\frac{\mathcal{R}^{\omega l}_\lambda}{\mathcal{I}^{\omega l}_\lambda}\right|^2\right).
\ee
Now, the incident energy density for a unit amplitude EM wave turns out to be $2\omega^2$ (see \cite{Cardoso:2019dte} for scalar case that will be multiplied by two for two polarization states of EM field). Finally, the total ACS reads as,
\be\label{acs.def}
\sigma=\frac{\pr_t\mathcal{E}}{2\omega^2}=\sum \sigma_l=\sum_{l, \lambda}\frac{\pi (2l+1)}{2}\left(1-\left|\frac{\mathcal{R}^{\omega l}_\lambda}{\mathcal{I}^{\omega l}_\lambda}\right|^2\right).
\ee
where $\sigma_l$ represents the partial ACS. Next, we will discuss the procedure to evaluate this quantity numerically.
\begin{figure*}[t]
\includegraphics[width=\linewidth]{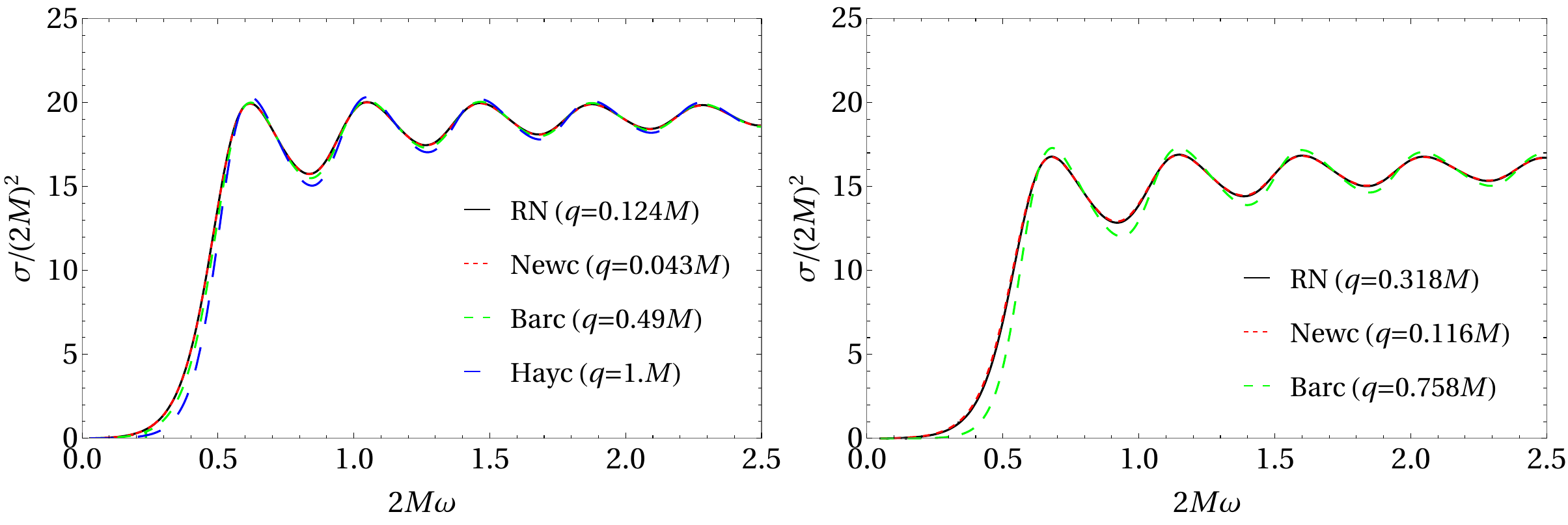}
\caption{Total ACS for the EM field with frequency ($\omega$): In the left panel, we have considered the Fan Wang generic class RBHs along with RN BH for different charges within the respective critical charges of the RBHs. In the right panel, we have excluded the Hayward BH to investigate the similarities more closely for RBHs with lower critical charges.}\label{RNNBHtotACS}
\end{figure*}
\section{Numerical evaluation of the absorption cross section (ACS)}\label{numeric.analysis}
In this section, we will present the numerical procedure to evaluate the ACS of the RBHs for the EM field. The first step is to solve the perturbation equation \eqref{chi1chi2.eom} with the ingoing initial condition set near the horizon, $r\to r_h(1+\epsilon)$, where $\epsilon<<1$ and $r_h$ represents the event horizon of the spacetime under consideration. We evaluate the solution at $r_\infty=150 M$, far from the event horizon. However, varying this evaluation point does not hamper the stability of the final results of our analysis. The next step is to match the solution with the ansatz form considered for $r\to \infty$ \eqref{bc.chi1chi2} and find out the reflection and incident coefficients. Then, by inserting these values for different modes ($l$) in the definition of ACS \eqref{acs.def}, one will be able to obtain the partial ACS of RBHs for the EM field. We verified that our numerical methodology accurately reproduces the existing results for the scalar field from \cite{Macedo:2014uga}, as well as for the electromagnetic field from \cite{Crispino:2007qw, deOliveira:2019tlk}.

In the left panel of Fig.\ref{NBHandNpartACS}, we have presented the partial ACS, $\sigma_l$, of RBHs in the Fan-Wang generic class. With a fixed charge, we observe that the peak of the ACS is higher for RBHs with higher critical charges, $q_{crit}$. In the right panel of Fig.\ref{NBHandNpartACS}, we specifically consider the New class RBH. Here, the peak decreases as we increase the charge $q$, restricted within $q_{crit}$. For the Simpson-Visser class, the partial ACS is plotted in Fig.\ref{SV.partial.ACS}, which exhibits a decreasing behaviour with the increment of the regularity parameter $r_0$. Importantly, one can deduce that the nature of these partial ACS is in accordance with the characteristics of the effective potential, $V_{eff}$, discussed before.

To find out the total ACS, we have evaluated the sum \eqref{acs.def}, considering $l$ up to $l=6$. In the left panel of Fig.\ref{NBHandNtotACS}, we have presented the total ACS of RBHs for the EM field in the Fan-Wang generic class. Its overall behaviour reflects the nature of partial ACS and it oscillates in the high frequency region. In the right panel of Fig.\ref{NBHandNtotACS}, we specifically consider the New class RBH and show the behaviour for different charges, $q$, restricted within $q_{crit}$. For the Simpson-Visser class, the total ACS has been plotted in Fig.\ref{SV_tot_acs_diffr0}, by varying the regularity parameter $r_0$.

Coming to the scenario where there might be a potential resemblance between the Fan-Wang generic class RBHs in NLED theory and the Reissner-Nordström (RN) BH in linear theory, we have analyzed the total ACS for various charges, as shown in Fig. \ref{RNNBHtotACS}. In the left panel, it is evident that the new class of RBHs closely resembles the RN BH \cite{Crispino:2008zz} across the entire frequency range for the given values of the charge, $q$, for respective classes. However, deviations begin to appear as we move from the new class to Bardeen and then to Hayward. This difference is more pronounced in the low-frequency region, but from mid to high frequencies, they converge towards the RN results. The choice of charges is not unique, by considering low to moderate values for the RN BH, similarities can be explored within the RBHs with different charges. Total ACS for another such choice of the set of charges has been provided in the right panel of Fig.\ref{RNNBHtotACS} excluding the Hayward class. We should mention here that very similar analyses for the scalar field have been worked out in the case of Bardeen class \cite{Macedo:2014uga}, Hayward \cite{dePaula:2023muc}. However, for the EM field, in \cite{Lopez:2022uie} the authors have mentioned these correspondences with RN BH but have not been studied rigorously. Therefore, our analysis of the EM field in the context of Bardeen and Hayward and specifically, for the new class of RBHs are comparatively new.

\section{Conclusion and future outlook}\label{concl}
We have analyzed the ACS for the EM field in static spherically symmetric RBH models, with the aim of comparing these models and examining how the ACS depends on the parameters of the RBHs. To perform the computation, we introduced a gauge invariant approach bypassing the procedure of gauge fixing for the EM field and followed that throughout our whole analysis. We find this approach to be effective and easier to follow. However, the applicability of the present methodology to the case of rotating BH spacetimes \cite{Leite:2018mon} should be investigated further.    

In the first case, we have considered the generic Fan-Wang class, which constitutes, Hayward \cite{Hayward:2005gi}, Bardeen \cite{Bardeen1968}, and a new class of RBHs \cite{Fan:2016hvf}. We find that for orbital mode, $l=1$, the partial ACS increases for RBHs with higher critical charge in the low-frequency region. However, in the high-frequency region, this variation vanishes. For higher orbital modes, the non-zero partial ACS starts at a higher frequency, and, it reveals the same feature with lower amplitude. This is in accordance with the fact that higher orbital modes give rise to the higher potential barriers. For individual classes, the increasing charges also generate a higher potential barrier, correspondingly, the partial ACS will be lower. This particular feature is also shared by the charged BH spacetime \cite{Crispino:2008zz}. All these characteristics are reflected in the total ACS of the respective RBHs, along with the oscillating feature in the high-frequency region. Regarding the possible correspondence between the RBHs in NLED theory and RN BH in linear theory, we have found that the new class of RBHs closely resembles the RN BH for several choices of the charge. This conclusion can be understood from the behaviour of the Lagrangian in the weak field limit as has been discussed in Sec.\ref{diff.RBHs}. Furthermore, between the Bardeen and Hayward class RBHs, we have seen that the first one is closer to the case of RN BH than the latter. The current analysis of the EM field, combined with previous studies on the scalar field \cite{Macedo:2014uga, dePaula:2023muc}, leads us to argue that in certain scenarios, RBHs in NLED theory lose their nonlinear nature and behave like a linear theory, similar to the RN spacetime.

In the second case, we have considered the Simpson-Visser class RBHs. From our analysis, we find that the regularization parameter $r_0$ plays the same role as the charge in the Fan-Wang generic class. Increasing this parameter leads to a higher potential barrier, and naturally, the partial ACS becomes lower. Hence, the total ACS also decreases as we increase the value of $r_0$, in the low-frequency region. In the high frequency, it exhibits the same decreasing behaviour along with oscillations. Importantly, this particular spacetime reduces to that of the Schwarzschild BH when $r_0=0$ and is sometimes referred to as the Schwarzschild-Simpson-Visser class, a modified version of the Schwarzschild BH. Therefore, our analysis has been compared with that of the Schwarzschild spacetime also. However, the dependence of the ACS on the parameter $r_0$ requires further investigation and a deeper understanding of the underlying source of this class of RBHs \cite{Bronnikov:2021uta}. Our primary motivation for studying the uncharged Simpson-Visser class has been to check whether our gauge-invariant approach is effective for the given metric structure. In this context, it is worth mentioning that our present methodology can be easily applied to the case of charged Simpson-Visser spacetime \cite{Bronnikov:2021uta} and it will be interesting to explore the correspondence with the spacetime of RN BH, as we have done in the case for Fan-wang generic class RBHs.

\noindent
\textbf{Acknowledgments:}
The author would like to thank Debaprasad Maity for various useful discussions and constant support while developing the basic methodologies for this work.

\end{document}